\documentclass[fleqn,usenatbib]{mnras}
\usepackage{newtxtext,newtxmath}
\usepackage[T1]{fontenc}
\usepackage{mathptmx} 
\usepackage{graphicx} 
\usepackage{amsmath}  
\usepackage{amssymb}  

\newcommand{\msun}{{\rm M}_\odot}
\newcommand{\msunyr}{{\rm M}_\odot\,{\rm yr}^{-1}}
\newcommand{\nh}{n_{\rm H}}
\newcommand{\cc}{{\rm cm^{-3}}}
\newcommand{\kms}{{\rm km\,s^{-1}}}
\newcommand{\vrad}{v_{\rm rad}}

\title
[Origin of Misalignments around protostar]
{Origin of Misalignments: Protostellar Jet, Outflow, Circumstellar Disc, and Magnetic Field}

\author[S. Hirano \& M. N. Machida]{
Shingo Hirano\thanks{E-mail: hirano.shingo.821@m.kyushu-u.ac.jp}
and
Masahiro N. Machida
\\
Department of Earth and Planetary Sciences, Faculty of Sciences, Kyushu University, Fukuoka, Fukuoka 819-0395, Japan
}

\date{Accepted 2019 March 11. Received 2019 January 16; in original form 2018 June 16}
\pubyear{2019}

\begin{document}
\label{firstpage}
\pagerange{\pageref{firstpage}--\pageref{lastpage}}
\maketitle

\begin{abstract}
Recent observations uncover various phenomena around the protostar such as misalignment between the outflow and magnetic field, precession of the jet, and time variability of the ejected clumps, whose origins are under debate.
We perform a three-dimensional resistive magnetohydrodynamics simulation of the protostar formation in a star-forming core whose rotation axis is tilted at an angle $45^\circ$ with respect to the initial magnetic field, in which the protostar is resolved with a spatial resolution of 0.01\,au.
In low-dense outer region, the prestellar core contracts along the magnetic field lines due to the flux freezing.
In high-dense inner region, on the other hand, the magnetic dissipation becomes efficient and weakens the magnetic effects when the gas number density exceeds about $10^{11}\,\cc$.
Then, the normal direction of the flattened disc is aligned with the angular momentum vector.
The outflow, jet, and protostellar ejection are driven from different scales of the circumstellar disc and spout in different directions normal to the warped disc.
These axes do not coincide with the global magnetic field direction and vary with time.
This study demonstrates that a couple of misalignment natures reported by observations can be simultaneously reproduced only by assuming the star-forming core rotating around a different direction from the magnetic field.
\end{abstract}

\begin{keywords}
MHD --
methods: numerical --
stars: formation --
stars: protostars --
stars: magnetic field --
stars: winds, outflows
\end{keywords}

\section{Introduction}
\label{sec:intro}

A detailed physical process of the protostellar accretion phase, when a newborn protostar is growing via accretion, is still open question.
The dominant physical process of the star formation is `gravitational contraction' \citep{shu77} but young stellar objects (YSOs) also show `outflows.'
The dense star-forming region cannot be observed via optical frequency but longer-wavelength radio/infrared observation opened the window.
The first evidence of the unanticipated upstream object from YSOs, moving in opposite directions from the protostar, was Herbig-Haro (HH) objects reported around 1950 \citep{herbig51,haro50}.
The following observations revealed that HH objects are related with bipolar outflows and collimated jets \citep[see review by][]{reipurth01}.
These flows play an important role in the star formation via the mass and angular momentum transports during the accretion phase which governs properties of the star.

Recent updates of the instruments (e.g. ALMA and SMA) permit to resolve the small-scale phenomena around the circumstellar disc and give some observational clues on open questions, for example, disc wind model as the driving mechanism of outflow \citep[e.g.][]{bjerkeli16,alves17,hirota17,lee17}.
The detailed observations also presented some peculiar features which cannot seem to be explained by the standard protostellar accretion scenario.
\citet{hull13,hull17} found that the magnetic fields (B-fields) in protostellar cores are not correlated with outflows but randomly aligned \citep[see also][]{chapman13,alves18}.
\cite{stephens17} also showed misalignments of outflows and B-fields which are perpendicular to the filamentary gas cloud, in which prestellar cores are embedded.
In addition, there are other phenomena which are not fully explained; e.g. knot episodically ejected from the protostar to different directions \citep[e.g.][]{riaz17}.
The observational facts seem to be inconsistent with the standard picture of the protostar formation that the B-field couples with the gas component during the protostellar core contraction and the B-field direction becomes parallel to outflows.

Theoretically, the generation and propagation mechanisms of protostellar outflow and jet are still under debate.
The magneto-hydrodynamic wind was firstly studied by \citet{blandford82} and many researchers after the work.
Numerical simulation is one way to investigate the complex phenomena in the star-forming core.
One of the physical candidates of the misalignment between the B-field and  protostellar outflows is turbulent motion \citep[e.g.][]{hennebelle11,seifried12,seifried13} but recent simulation suggests strong turbulence forbids forming the circumstellar disc and driving the disc winds \citep{lewis18}.
Another possible origin is the initial directional misalignment between the global B-field and rotation of the prestellar core.
The first MHD simulation of the non-parallel case \citep{matsumoto04} showed the complex phenomena around accreting protostar and followed by some works
\citep[e.g.][]{machida06,hennebelle09,ciardi10,joos12,matsumoto17}.
However, previous studies adopted the sink cell technique, with the sink radius $\sim 1$--$10$\,au, to compute long-term evolution.
The small-scale phenomena including the protostellar jet and magnetic dissipation process could not be reproduced in such studies.
On the other hand, some studies performed simulation with resolving the small-scale phenomena however they began the simulations from the accretion phase by assuming a newborn protostar embedded in an artificial structure of the collapsed core \citep[see reviews by][]{konigl00,pudritz07}.

The aim of this study is to develop a physical understanding of the misalignment nature around protostar.
We perform a three-dimensional (3D) resistive magnetohydrodynamics (MHD) simulation \citep[][firstly computed long-term jet evolution with resolving from the star-forming core to protostar]{machida14}, in which the rotation axis is tilted at an angle $45^\circ$ with respect to the initial B-field direction.
We calculate the gravitational collapse of a prestellar core and emergence of low-velocity outflow and jet.
We analyse the time evolution of axes for outflow, jet, disc normal, angular momentum, and B-field during the protostellar accretion phase.
We find that a couple of unexplained observational features are reproduced by assuming only the initial misalignment of the B-field direction and rotation axis of the prestellar core.

\section{Methodology}
\label{sec:method}

We perform a simulation of protostar formation from a magnetized prestellar core by using a 3D MHD code \citep{fukuda99} which adopts the nested grid method to calculate the gravitational collapse \citep{matsumoto03}.
To satisfy the Jeans condition in the collapsing region, the grid generation procedure is adopted in which the Jeans length is resolved, at least, $16$ cells.
The grid width is $64^3$ and the cell size is $740$\,au in the lowest level whereas $0.01$\,au in the highest level.
The code also solves resistive MHD equations to treat the magnetic dissipation \citep{machida07}.

\subsection{Initial condition}
\label{sec:method_ics}

The simulation starts from a rigid rotating prestellar core in a uniformly magnetized density background (interstellar medium).
The initial cloud has  a Bonner-Ebert density distribution with the central number density, $n_0 = 6.0 \times 10^5\,\cc$, and isothermal temperature, $T_0 = 10$\,K, in which we set the cloud radius as twice the critical Bonner-Ebert radius.
Before the calculation, the cloud density is enhanced by a factor of $1.8$ to promote the gravitational contraction.
Thus, the central density of the prestellar core is $n_{\rm 0,c}= 1.08\times  10^6\,\cc$, while the background density is $n_{\rm 0,bg} =  1.3 \times 10^4\,\cc$.
The mass and radius of the prestellar cloud is  $M_{\rm cl}=2.1\,\msun$ and  $R_{\rm cl}=1.18 \times 10^4$\,au, respectively.
A rigid rotation with $\Omega_0 = 1.4 \times 10^{-13}\,{\rm s^{-1}}$ is imposed.
The core is magnetized as $B_0 = 5.73 \times 10^{-5}$\,G.
The ratios of thermal, rotational, and magnetic to gravitational energy of the initial core are $\alpha = 0.390$, $\beta = 0.026$, and $\gamma = 0.495$, respectively.

Unlike the previous simulation \citep{machida14} in which the rotational axis of the core is parallel to the direction of the B-field ($0^\circ$), these two directions are inclined by $45^\circ$ in this study.
We set that the initial direction of the B-field ($B_0$) is in $Z$-axis positive direction whereas the rotational axis ($J_0$) is in $45^\circ$ tilted direction between $Z$- and $X$-axes.

We set the protostar formation epoch ($t = 0$) as when the collapsing centre reaches $\nh = 10^{18}\,\cc$ and proceed the simulation of the protostellar accretion phase for the first $300$\,yr.
We define the protostar as a region where the number density exceeds  $\nh \ge 10^{18}\,\cc$ to derive the protostellar mass.

\subsection{Measurements of axes}
\label{sec:axes}

We analyse the simulation results to measure the time evolution of four axes defined as follows:
(a) directions of the outflows with different radial velocities,
\begin{equation}
	\textbf{\textit{v$_{\rm out}$}}(v_{\rm min}, v_{\max}) = \frac{1}{V(v_{\rm rad})} \int_{v_{\rm min} \le v_{\rm rad} \le v_{\rm max}} \textbf{\textit{v}} \, {\rm sign}(z) dV\,,
    \label{eq:v}
\end{equation}
where $V(v_{\rm rad})$ denotes a volume where the radial velocity is within a range between $\{v_{\rm min}, v_{\rm max}\} = \{1, 5\}$, $\{10, 20\}$, and $\{20, 30\}\,\kms$, and \textbf{\textit{v}} is the fluid velocity at each cell, (b) normal vectors of the flattened disc, $\textbf{\textit{n}}_{\rm disc}(n_{\rm th})$, corresponding to the eigenvector of the inertial tensor\footnote{The other inertial moments in Equation~(\ref{eq:n1}) can be obtained by permuting the indices ($x$, $y$, and $z$) of Equations~(\ref{eq:n2}) and (\ref{eq:n3}).} \citep{machida06,matsumoto17},
\begin{eqnarray}
	\textbf{\textit{I}} &=& \left(
		\begin{array}{ccc}
		 I_{\rm x}  & -I_{\rm xy} & -I_{\rm xz} \\
        -I_{\rm yx} &  I_{\rm y}  & -I_{\rm yz} \\
		-I_{\rm zx} & -I_{\rm zy} &  I_{\rm z}
		\end{array}
    \right)\,,\label{eq:n1}\\
    I_{\rm x}  &\equiv& \int_{n \ge n_{\rm th}} (y^2 + z^2) \rho dV\,,\label{eq:n2}\\
	I_{\rm xy} &\equiv& \int_{n \ge n_{\rm th}} (xy) \rho dV\,,\label{eq:n3}
\end{eqnarray}
associated with the smallest eigenvalue for three threshold densities $n_{\rm th} = 10^7$, $10^{10}$, and $10^{13}\,\cc$, (c) rotational axes (angular momentum vectors),
\begin{equation}
	\textbf{\textit{J}}_{\rm ang}(n_{\rm th}) = \frac{1}{M(n_{\rm th})} \int_{n \ge n_{\rm th}} (\textbf{\textit{r}} \times \textbf{\textit{v}}) \rho dV\,,
    \label{eq:J}
\end{equation}
where $M(n_{\rm th})$ is the total mass within $n_{\rm th}$, and (d) direction of the magnetic-field averaged by volume at different scales,
\begin{equation}
	\textbf{\textit{B}}(r_{\rm th}) = \frac{1}{V(r_{\rm th})} \int_{r \le r_{\rm th}} \textbf{\textit{B}}(r) dV\,,
    \label{eq:B}
\end{equation}
where $V(r_{\rm th})$ is a spherical volume with radius $r_{\rm th} = 10$, $30$, $100$, $10^3$, and $10^4$\,au.

\section{Results}
\label{sec:result}

We began with the calculation from the gas collapsing phase prior to protostar formation, while, in this paper, we only showed the results during the gas accretion phase following protostar formation to focus on the structures of circumstellar disk and protostellar outflows.
The cloud evolution during the gas collapsing phase was investigated in many past studies \citep[e.g.][]{banerjee06,machida06,tomida15,tsukamoto15,wurster18}.

At first, we overview the driving of outflows in the accreting cloud.
During the gas accretion phase, outflows with different velocities emerge and sweep the surrounding gas.
The circumstellar plasma is accelerated by the magneto-centrifugal force and/or magnetic pressure driven in the strong or weak B-field regions, separated by so-called `dead zone' at $n_{\rm H} > 10^{11}\,\cc$ where the magnetic dissipation becomes efficient and weakens the B-field strength.\footnote{The threshold density is determined by the local chemical properties and becomes the same value independent on the initial orientation of the B-field \citep[see fig.~1 in][]{machida07}.}
These two regions drive two different flows, as shown in previous studies \citep[e.g.][]{machida07,tomida13,tsukamoto15}.
First, the slow outflow (hereafter, the low-velocity outflow) with a wide opening angle is driven from the outer disc region due to the magneto-centrifugal force \citep{blandford82}.
Second, the fast-collimated jet (hereafter, the jet) appears from the inner disc region and is accelerated mainly by the magnetic pressure \citep{lynden-bell03}.
In addition, episodically knotty ejection (hereafter, the protostellar ejection or knots) is driven by the magnetic pressure in response to the intermittent mass accretion from the inner-edge of the disc onto the protostar (see Movie~1).
Their driving mechanisms are the same as in the alignment case \citep{machida08,machida14}.
However, since the disk is warped with different scales, the direction of each outflow changes.
This is natural consequence of misalignment between magnetic field and rotation axis both of which forms anisotropic structure or disk.

\begin{figure}
\begin{center}
\includegraphics[width=1.0\columnwidth]{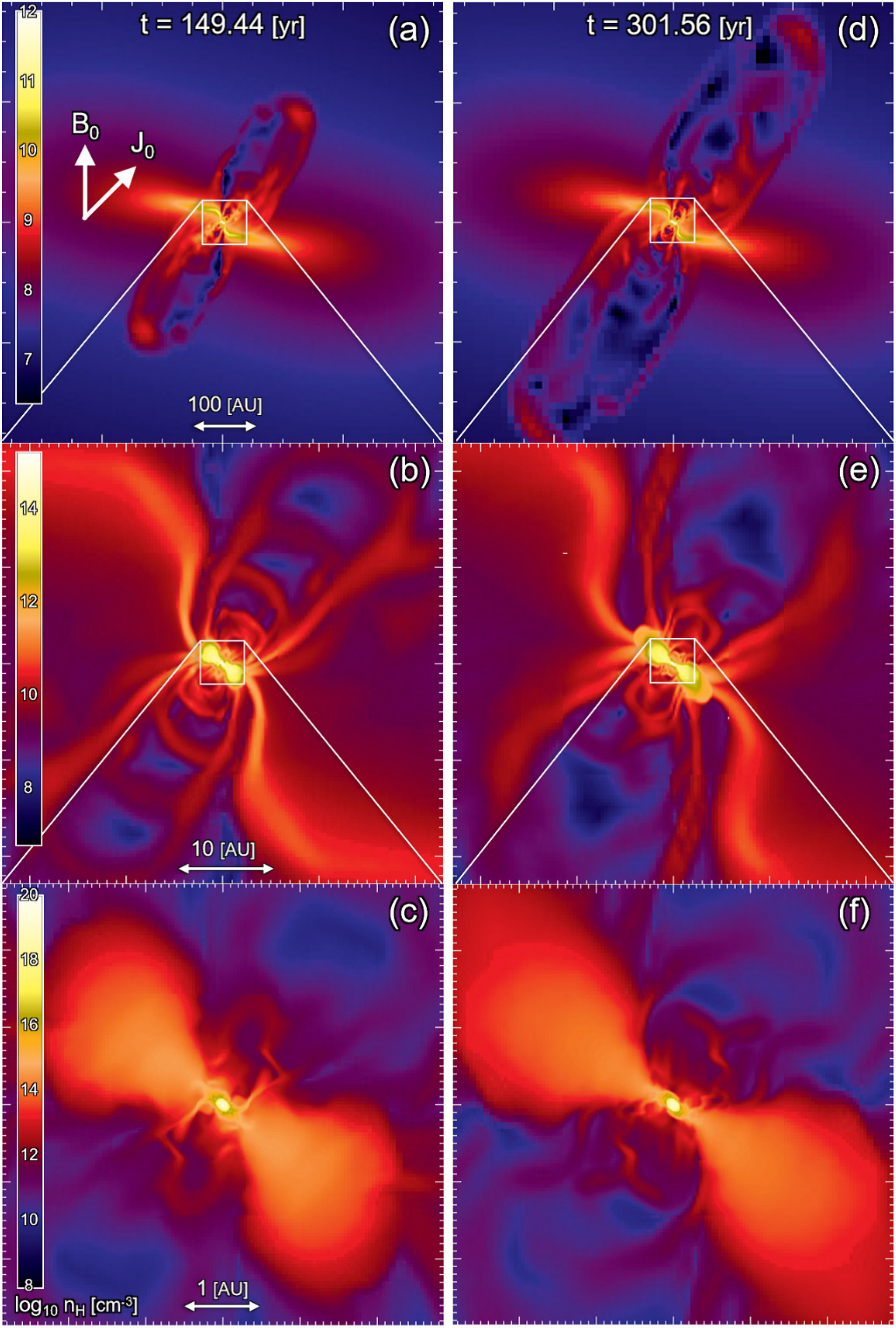}
\end{center}
\caption{
Density distributions on the $y = 0$ plane with box sizes of $739$, $46$, and $5.8$\,au (top, middle, and bottom rows) at $t = 149.44$  and $301.56$\,yr after the protostar formation (left and right columns).
There are three different flows: low-velocity outflow from the outer disc (top), jet from the inner disc (middle), and protostellar ejection from the accreting protostar (bottom).
The driving regions of low-velocity outflow and jet are divided by the dead-zone in the disc with $\nh \sim 10^{11}\,\cc$ where the magnetic dissipation becomes efficient. (see Movie~1, available online)
}
\label{f1}
\end{figure}

\subsection{Misalignments between circumstellar properties}

Figure~\ref{f1} plots the density distributions around the accreting protostar at three different scales, in which two different epochs of $t=149.44$ and $301.56$\,yr  are shown.
Figures~\ref{f1}(a) and (d) indicate that the outflows gradually evolve with time.
The shell-like structures seen in Figs.~\ref{f1}(b) and (e) are caused by intermittently driven high-velocity flows or jets (for details, see below).
The central yellow region in Figs.~\ref{f1}(c) and (f) corresponds to the protostar that is enclosed by the warped disk-like structure.
Figure~\ref{f1} indicates that although the disk and outflow system appears around protostar as usually seen in previous studies \citep[e.g.][]{tomisaka02,banerjee06,hennebelle08a,hennebelle08b,machida14}, there exist a complex nature of circumstellar environment caused by the initial directional misalignment between the global B-field and rotation axis of the prestellar core.

\begin{figure}
\begin{center}
\includegraphics[width=1.0\columnwidth]{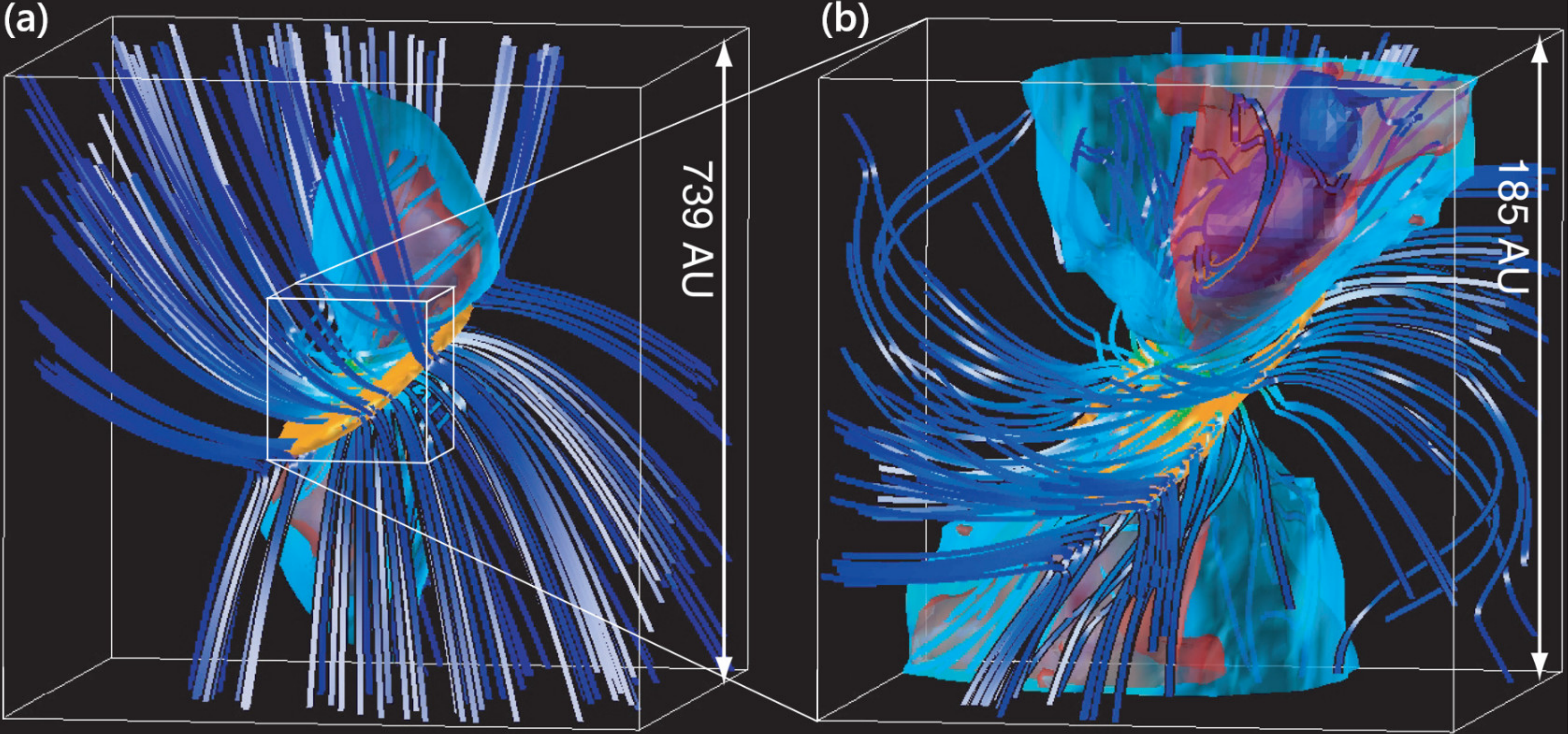}
\end{center}
\caption{
Three-dimensional structure at $t = 301.56$\,yr after the protostar formation with box sizes of ($a$) $739$ and ($b$) $185$\,au.
The orange iso-density surfaces indicate a pseudo-disc ($\nh = 5 \times 10^9\,\cc$; $a$) and rotationally-supported disc ($\nh = 10^{10}\,\cc$; $b$).
The tubes show the B-field lines, which is roughly aligned with the initial direction at large scale but twisted at small scale because of the coupling to the accreting gas.
The aqua, red, and blue iso-velocity surfaces at $\vrad = 8$, $15$, and $20\,\kms$ depict three outflows (low-velocity outflow, jet, and protostellar ejection). (see Movie~2, available online)
}
\label{f2}
\end{figure}

\begin{figure}
\begin{center}
\includegraphics[width=1.0\columnwidth]{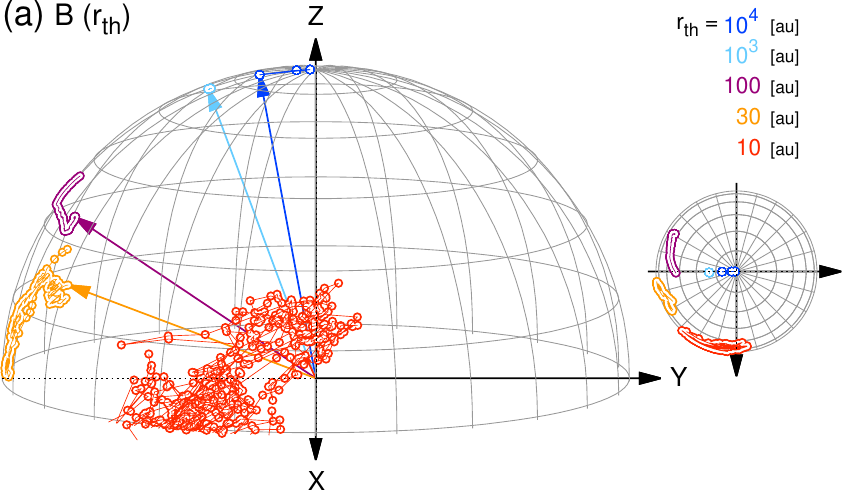}\\
\includegraphics[width=1.0\columnwidth]{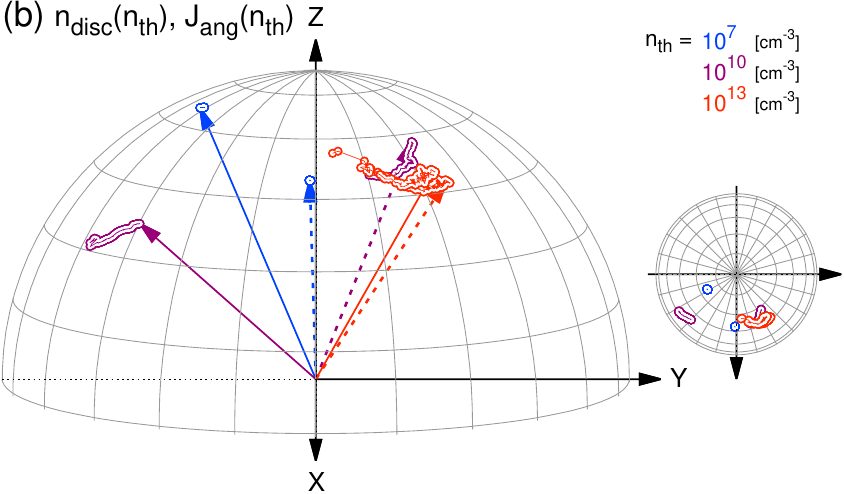}\\
\includegraphics[width=1.0\columnwidth]{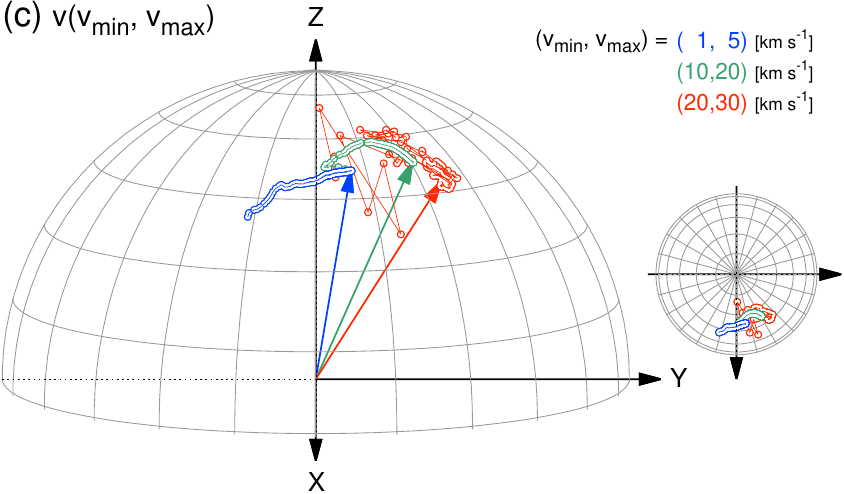}\\
\end{center}
\caption{
Time evolution of four directions:
($a$) B-field at five threshold radii $r_{\rm th} = 10$, $30$, $100$, $10^3$, and $10^4$\,au,
($b$) disc normal (solid arrows) and angular momentum (dashed) at three threshold number densities $n_{\rm th} = 10^7$, $10^{10}$, and $10^{13}\,\cc$, and
($c$) outflows with \{$v_{\rm min}$, $v_{\rm max}$\} = \{$1$, $5$\}, \{$10$, $20$\}, and \{$20$, $30$\}$\,\kms$.
The arrows indicate the final directions when we stop the simulation.
The initial direction of the B-field ($B_0$) is in $Z$-axis positive direction.
Whereas the rotational axis of the initial state ($J_0$) is tilted at $45^\circ$ from $Z$-axis to $X$-axis.
}
\label{f3}
\end{figure}

Figure~\ref{f2} shows three-dimensional views at $t=301.56$\,yr with different spatial scales, in which the viewing angle is the same between the panels.
Note that, in Fig.~\ref{f2}, we adjusted the viewing angle in order to exaggerate the angle differences in outflows.
Thus, the viewing angle in Fig.~\ref{f2} is not the same as that in Fig.~\ref{f1}.
From the figure, we can confirm that the direction of magnetic field lines, disk-like structure and outflows vary as the spatial scale differs.
Especially, the configuration of magnetic fields varies significantly with a slight change of the spatial scale.

To visually confirm the direction of each object, in Fig.~\ref{f3}, the directions of magnetic field (Fig.~\ref{f3}a), disk normal and angular momentum (Fig.~\ref{f3}b), and outflows (Fig.~\ref{f3}c) are plotted on the hemisphere.
Each direction of objects is estimated according to the procedure described in Section~\ref{sec:axes}.
Figures~\ref{f3}(a) and (b) show that the disk normal $n_{\rm disk}$ in the low-dense outer region ($n_{\rm thr}=10^7\cc$) is roughly parallel to the direction of the large-scale B-field  ($r_{\rm th}=10^4$ and $10^3$\,au).
This indicates that the disk-like structure, which corresponds to the pseudo-disk, is formed by the Lorentz force in a large scale.
The disk normal in the high-dense inner region ($n_{\rm th}=10^{13}\cc$) considerably differs from the directions of  B-fields, while it well agrees with the direction of the angular momentum (Figs.~\ref{f3}a and b), indicating that, in a small scale, the disk is formed mainly by the centrifugal force.
The directions of outflows also roughly coincide with the disk normal direction in the high-dense inner region (Figs.~\ref{f3}b and c).

Because outflows (low-velocity outflow, jet, and knots) are launched to the normal directions of the warped disc at the corresponding scales, their ejecting directions are not parallel to the initial magnetic field ($\textbf{\textit{B}}_0$).
Furthermore, directions of outflows are misaligned each other since the ejecting direction gradually changes depending on the launching scale (Fig.~\ref{f3}c).
The combination of these directional misalignments results in the hierarchical structure around in the star-forming core (Fig.~\ref{f1}).

\begin{figure}
\begin{center}
\includegraphics[width=1.0\columnwidth]{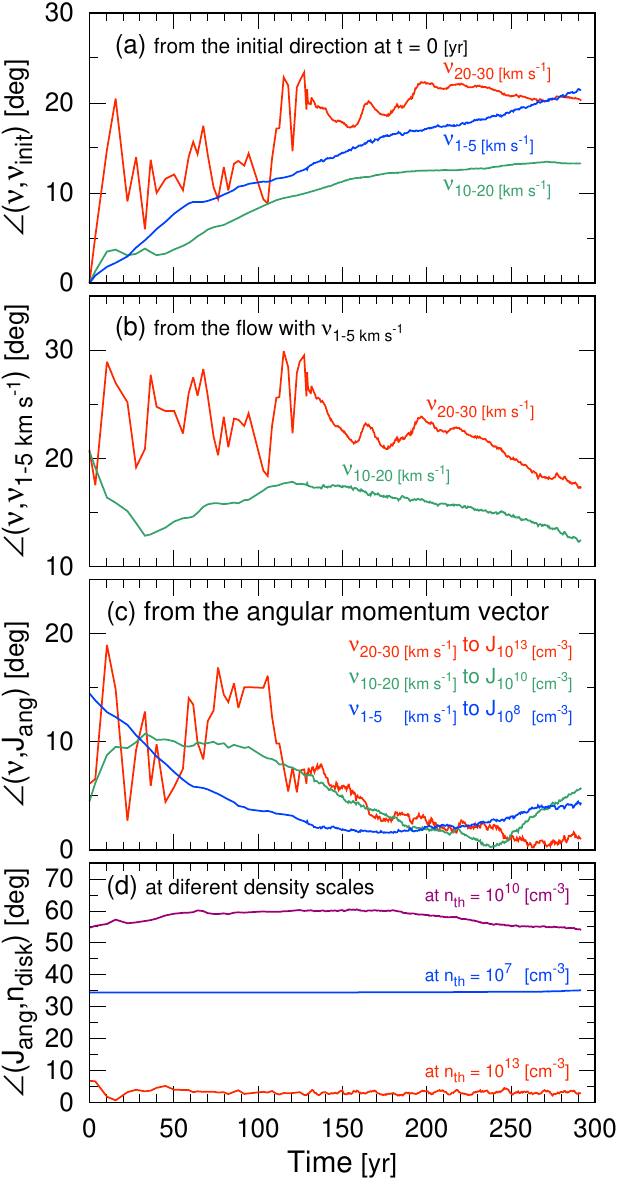}
\end{center}
\caption{
Time evolution for the first $300$\,yr from the protostar formation:
($a$) angles of flow directions (\textbf{\textit{v}}; Equation~\ref{eq:v}) compared to the initial orientation at the protostar formation for $\vrad = 1$--$5\,\kms$, $10$--$20\,\kms$, and $20$--$30\,\kms$,
($b$) angles of flow directions for $\vrad = 10$--$20\,\kms$ and $20$--$30\,\kms$ compared to one for $\vrad = 1$--$5\,\kms$,
($c$) angles between flow directions for three velocity ranges and rotational axes ($\textbf{\textit{J}}_{\rm ang}$;  Equation~\ref{eq:J}) for threshold densities $n_{\rm th} = 10^8$, $10^{10}$, and $10^{13}\,\cc$ at which the relative angles become minimum, and
($d$) angles between rotational axes and normal vectors to the discs ($\textbf{\textit{n}}_{\rm disc}$;  Equation~\ref{eq:n1}) for threshold densities $n_{\rm th} = 10^7$, $10^{10}$, and $10^{13}\,\cc$.
}
\label{f4}
\end{figure}

To quantify the difference in the angles of various objects, the angle differences are plotted  in Fig.~\ref{f4}.
Figure~\ref{f4}(a) shows the time evolution of the launch direction of flows with different velocity range relative to the initial orientation when the protostar appears.
In this study, we computed the launch directions (Equation~\ref{eq:v}) of low-velocity outflow, jet, and knots for flows with $\vrad = 1$--$5\,\kms$, $10$--$20\,\kms$, and $20$--$30\,\kms$.
Their ejecting directions shift by about $15$--$20$ degrees during $300$\,yr (Fig.~\ref{f4}a).
In addition, the directional misalignment between flows decreases with time (Fig.~\ref{f4}b).
The relative angle between the low-velocity outflow and jet, $\angle(\textbf{\textit{v}}_{1-5\,\kms}, \textbf{\textit{v}}_{10-20\,\kms})$, monotonically decreases after $150$\,yr of the protostar formation from $17.5^\circ$ to $12.3^\circ$ (with $0.0347^\circ$\,yr$^{-1}$).
The angle between the low-velocity outflow and knots, $\angle(\textbf{\textit{v}}_{1-5\,\kms}, \textbf{\textit{v}}_{20-30\,\kms})$, starts to decrease after $200$\,yr and changes from $24^\circ$ to $17^\circ$ (with $0.07^\circ$\,yr$^{-1}$).
This alignment tendency indicates that the warped structure of the circumstellar disc, whose normal direction determines the launching orientation, is eliminated with time at every scale.
Thus, the misaligned nature of outflows can be seen in a very early phase of star formation.
However, further time integration is necessary to verify this trend \citep[see also][]{ciardi10}.

Figure~\ref{f4}(c) plots the angles between the direction of outflows  and  rotational axis.
In the figure, although a slight difference (blue) and oscillation (red) between the angles can be seen in the early phase ($t<100$\,yr), the angle differences  become very small for $t>100$\,yr, indicating  that the direction of outflows are controlled by the angular momentum.

At last, we describe the angle difference between rotation axis and disk normal.
As shown in Fig.~\ref{f4}(d), the disk normal is not aligned with the rotation axis in the low-dense outer region ($n_{\rm th}=10^7$, $10^{10}\cc$), while the disk normal is well aligned  with the rotation axis in the high-dense inner region ($n_{\rm th}=10^{13}\cc$).
In the prestellar cloud, the magnetic energy is larger than the rotational energy.
Thus, in the low-dense outer region, the Lorentz force produces a disk-like structure or pseudo-disk.
On the other hand, the magnetic field dissipates in the range of $10^{11}\cc \lesssim n \lesssim 10^{15}\cc$ where magnetic field  becomes considerably weak \citep{nakano02} and the rotation or centrifugally force produces a disk.
As a result, in the high-dense inner region ($n_{\rm th}=10^{13}\cc$), the disk normal coincides with the rotation axis.

\subsection{Protostellar growth and mass accretion and ejection rates}

\begin{figure}
\begin{center}
\includegraphics[width=1.0\columnwidth]{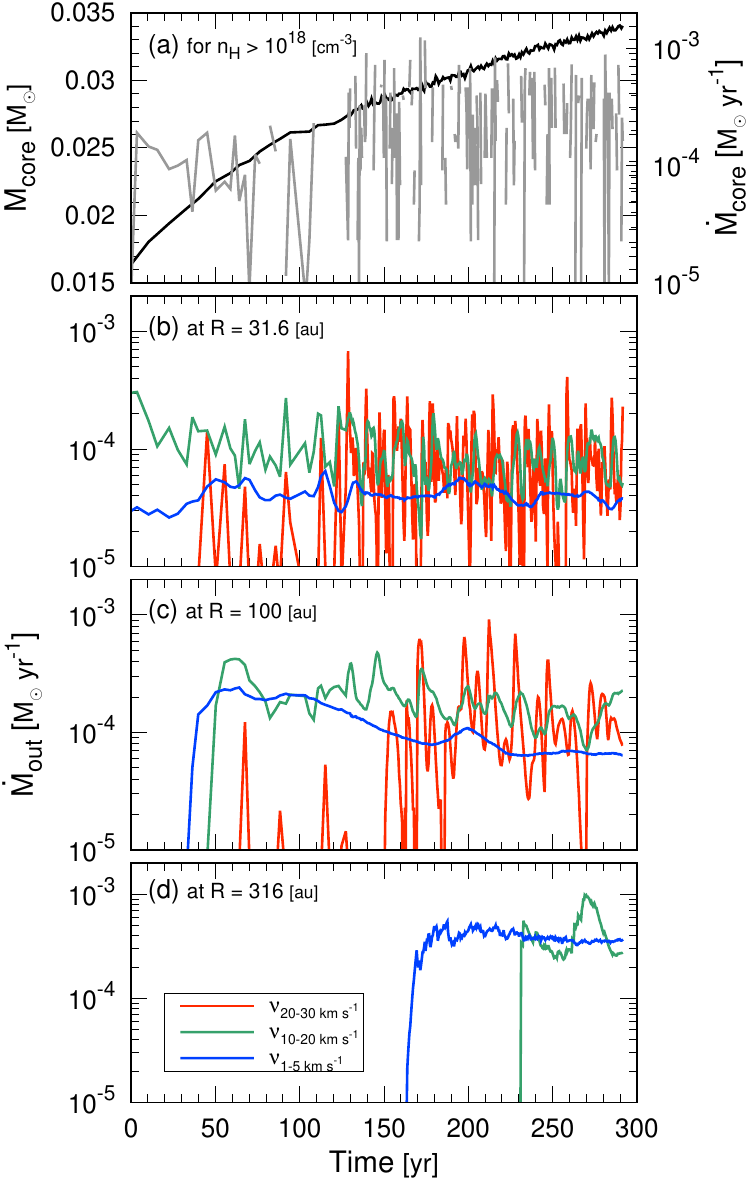}
\end{center}
\caption{
Time evolution for the first $300$\,yr from the protostar formation:
($a$) central core mass (black) and mass accretion rate onto the core (grey),
($b$--$d$) mass ejection rates with different velocity flows ($\vrad = 1$--$5\,\kms$, $10$--$20\,\kms$, and $20$--$30\,\kms$) at radius $R = 31.6$, $100$, and $316$\,au from the central protostar.
}
\label{f5}
\end{figure}

This study calculated the first $300$\,yr of the mass accretion phase during which the protostar grows up to $0.034\,\msun$.
The protostellar mass and mass accretion rate are plotted against the elapsed time after protostar formation in Fig.~\ref{f5}(a).
The figure indicates that the episodic  accretion occurs, in which the mass accretion rate varies from $\sim10^{-5}\msunyr$ to $\sim 10^{-3}\msunyr$  with the average of $6.05 \times 10^{-5}\,\msunyr$.
The averaged mass accretion rate agrees well with that derived in self-similar solutions \citep{larson03}.

The mass ejection rates in different spatial scales are plotted in Figs.~\ref{f5}(b)--(d).
The ejection rates are estimated with different velocity thresholds of $\vrad = 1$--$5\,\kms$, $10$--$20\,\kms$, and $20$--$30\,\kms$.
The mass ejection rate does not show significant change in a large scale (Fig.~\ref{f5}d).
Especially, the low-velocity outflow ($\vrad = 1$--$5\,\kms$) has an almost constant mass ejection rate of $\sim3.5 \times 10^{-4} \msunyr$.

On the other hand, the high-velocity components (jet and knots; $\vrad = 10-20\,\kms$, and $20-30\,\kms$) in the small scales have a periodic time variability (Figs.~\ref{f4}b and c).
The period of time variation of mass ejection in the small scale (Fig.~\ref{f5}b) is within $< 10$\,yr which roughly corresponds to that of  mass accretion.
The period of time variation of mass ejection in the middle scale (Fig.~\ref{f5}c) becomes longer than that in the small scale (Fig.~\ref{f5}b).
Thus, Fig.~\ref{f5} indicates that the mass accretion is more synchronised with the mass ejection rate in the inner disk region or small scale  than that in the middle and large scales.
Thus, it is expected that the eruption of knots at the innermost region is caused by the episodic accretion (Figs.~\ref{f5}a and b).
The jet and knots form shell like structures inside the low-velocity outflow, which are not centred but deflected (Fig.~\ref{f1}).
The ejected clump mass is estimated as $\sim\!10^{-4}\,\msunyr \times 10\,{\rm yr} \simeq 10^{-3}\,\msun$ (Fig.~\ref{f5}c).

\section{Interpretation of observations}
\label{sec:discus}

This study first presents the emergence of low-velocity outflow, jet, and knot during the early accretion phase of Class 0 object formed from the prestellar core whose rotational axis is not parallel to the initial B-field direction.
The simulation result suggests a couple of phenomena which were not found in the parallel case: scale- and time-dependent misalignments between the directions of B-field and outflows (Section~\ref{sec:result}).
These phenomena are possible to explain unexplained features reported by recent observations (refereed below; see also Section~\ref{sec:intro}) or be found by the future observations.
We summarise the phenomena during the star formation obtained in this simulation with references of the corresponding observations.
\begin{enumerate}

\item
Scale dependence of the B-field distribution.
At large scale, the direction of B-field averaged by volume  (Equation~\ref{eq:B}) is almost parallel to the initial orientation \citep[Fig.~\ref{f2}a; cf.][]{girart06}.
At small scale, on the other hand, the B-field lines are dragged by the contracting gas rotating around a certain axis \citep[Fig.~\ref{f2}b; cf.][]{galametz18}.
At the innermost region, the B-field on the circumstellar disc shows a radial configuration \citep[Figs.~\ref{f3}a and \ref{f2}; cf.][]{alves18}.\footnote{The winding the B-field in a spiral configuration on the innermost disc is occurred in the parallel case, independent of the initial setting of the rotational axis.}
As also seen in Fig.~\ref{f3}, in a small scale ($\lesssim 10$--$100$\,au), the magnetic vectors are almost randomly distributed.
Thus, various configurations of the magnetic field are expected in the small scale.
We will show various configuration of magnetic field in a forthcoming paper.

\item
Warped disk structure.
During the prestellar core collapse, the disk normal (Equations~\ref{eq:n1}--\ref{eq:n3}) firstly aligns with the B-field direction at low-dense region but finally with the angular momentum vector at high-dense region (Fig.~\ref{f3}).
Such the distortion of the disc structure and rotational axis will be certified by the detailed observation \citep[cf.][]{mayama18,sakai19}.

\item
Low-velocity outflow, jet, and knots are driven by different radii.
Since such flows are driven along the rotation axis (or disk normal) at each driving radius (Fig.~\ref{f4}c) and the disk normal differs in each scale, the flows are not aligned each other.
Thus, there exists different angles between flow directions \citep[Fig.~\ref{f3}c; cf.][]{matsushita19}.
In addition, the flow directions are not matched with the direction of the global B-field, which can be already seen in observations \citep[cf.][]{hull17}.
The directional misalignments among outflows decrease with time (Fig.~\ref{f4}b) then the directions are expected to tend to align in the later phase at closer to the protostar.

\item
The directions of B-field, disk normal, and outflows are dynamically changed as the spatial scale differs (Fig.~\ref{f3}), indicating that the classical simple picture of star formation is not very useful, at least, for considering the directions of objects.
The protostar ejection occurs episodically due to the intermittent mass accretion (Figs.~\ref{f5}a and b) and the ejecting direction changes with time \citep[Figs.~\ref{f3}c and \ref{f4}a; cf.][]{takami11,riaz17}.
The gradual change of the outflow directions, about $15$--$20^\circ$ during the first $300$\,yr, enhances the sweeping volume by the flows (see Movie~2).

\end{enumerate}
Other mechanisms are also possible to originate the above feature, e.g. the impact of the turbulence on the misaligned B-fields \cite[e.g.][]{seifried15}.
This work shows that a couple of phenomena listed above are simultaneously reproduced by considering only the non-parallel configuration of the prestellar core without hypothesizing other mechanisms.
To obtain the realistic picture which is observed, of course, we have to perform a synthetic observation using the simulation result, but it is beyond the scope of this study.

\section{Discussion}
\label{sec:discussion}

In this study, we calculated the cloud evolution resolving protostar without using sink cell technique.
In many studies, the sink cell technique is used to accelerate the time integration of simulation at the expense of resolving small scale structures, in which the region around the protostar within $\lesssim 1$--$10$\,au is masked by the sink cells and not resolved \citep[see references in][]{li14}.
The simulations with sink cells can follow the long-term evolution of circumstellar disk in a later stage.
Instead, with sink cells, the researchers cannot confirm the small-scale structure near the protostars, where a nascent disk appears \citep{bate98,machida10} and the low-velocity outflow and jet driving begins to occur \citep{tomisaka02}.
Moreover, a very early stage of star formation cannot be investigated by such simulations, because both very young disk and outflows, which have very small sizes, cannot be resolved.
On the other hand, since we did not use sink cells in this study, we can resolve the small-scale structure around a very young protostar.
Thus, we can investigate both the disk forming and  jet driving regions during a very early evolutionary stage.

This study can also constrain the driving mechanism of outflow and jet.
There exist controversial scenarios for the low-velocity flows: entrainment by the high-speed jets \citep{arce07} and direct driven from the disk outer edge \citep{tomisaka02,hennebelle08a}.
As described in Section~\ref{sec:intro}, although recent ALMA observation supported the latter (direct driven scenario), further observations are necessary to determine their driving mechanism reliably.
If the angle difference between low- and high-velocity flows are usually observed around very young protostars, our result can strongly constrain the flow driving mechanism.
Currently, we are observing the very early stage of star formation by ALMA.
The Keplerian disks around very young protostars were confirmed by some researchers \citep{okoda18,lee18}.
In addition, the misalignment between low- and high-velocity outflows was recently observed \citep{matsushita19}.
Now, the spatial resolution of observations of nearby star-forming regions is comparable to or exceeds that of past star formation simulations \citep{li14}.
In addition to high-spatial resolution simulations, more observations of very early stage of the star formation are necessary to correctly understand the star formation process.

At last, we comment on other non-ideal MHD effects.
We only considered the Ohmic dissipation as the non-ideal MHD effects, and ignored the other non-ideal MHD effects of ambipolar diffusion and Hall effect.
\citet{nakano02} showed that the Ohmic dissipation is primary mechanism to remove the magnetic flux in the collapsing cloud.
However, both ambipolar diffusion and Hall effect would influence the outflow driving and disk formation \citep{hennebelle16,marchand16}.
We need to calculate the accretion phase of star formation including all non-ideal MHD effects in future studies.

\section{Conclusion and future perspective}

In this study, using our non-ideal MHD simulation code, we calculated the evolution of a star-forming cloud from pre-stellar core stage until $\sim\!300$\,yr after protostar formation, and showed a picture of the early star-formation stage.
In our previous studies \citep{machida08,machida11,machida14a,machida14}, we adopted an idealized setting (B-fields parallel to angular velocity vectors) to simply analyse and determine the flow driving and disk formation conditions.
However, the magnetic field lines would be not perfectly aligned with the rotation axis in observations \citep[e.g.][]{shinnaga12}.
Thus, the initial setting adopted in this study (misalignment between B-field and angular velocity vector) is more realistic.
Note that we did not impose turbulence in the prestellar cloud core, which would naturally reproduce the misaligned nature \citep{matsumoto17}, in order to simply compare this study with the aligned case \citep{machida14}.

\begin{figure}
\begin{center}
\includegraphics[width=1.0\columnwidth]{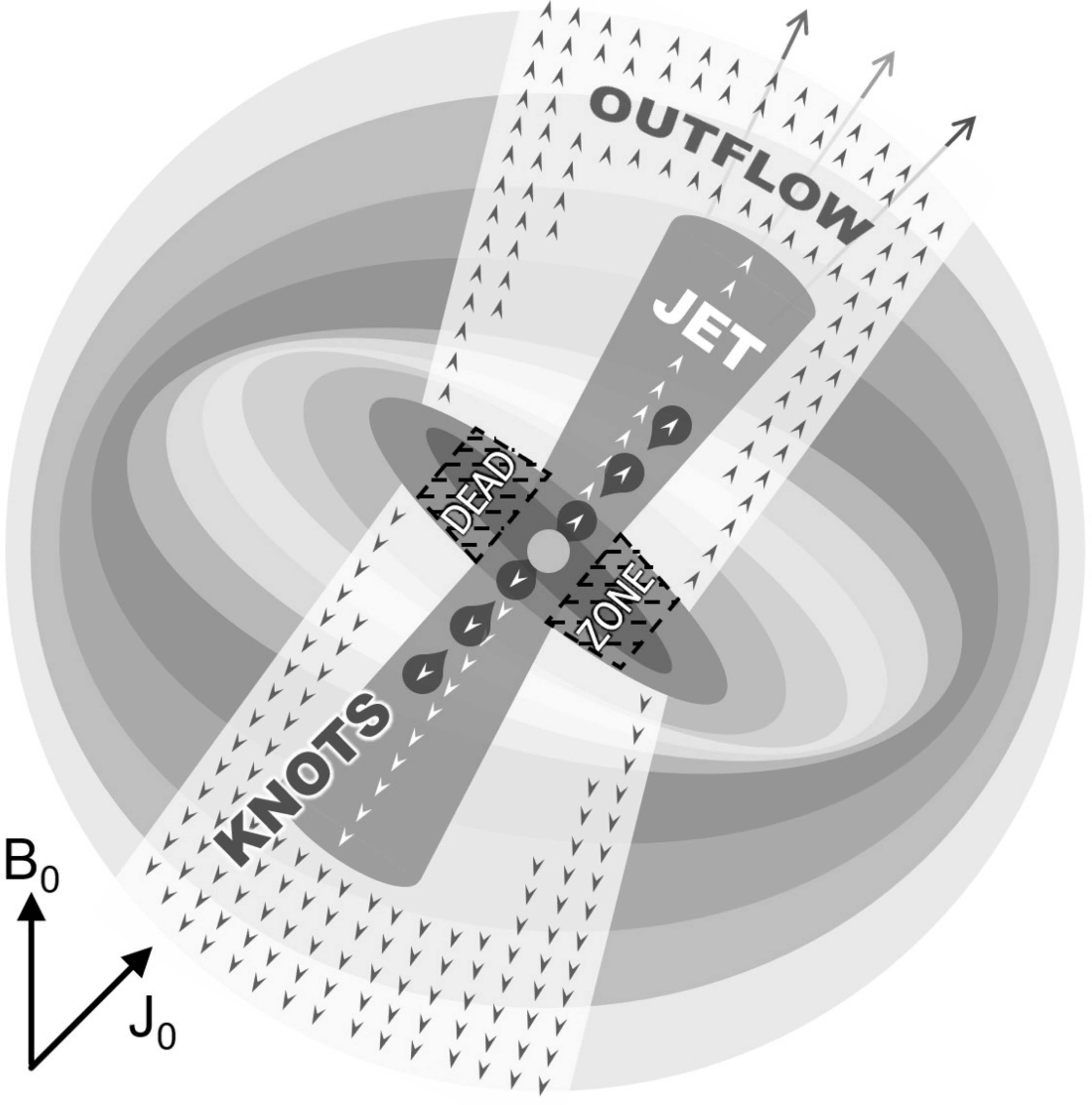}
\end{center}
\caption{
Schematic view of the simulation result.
The initial directions of the B-field ($\textbf{\textit{B}}_0$) and angular momentum vector ($\textbf{\textit{J}}_0$) are inclined at $45^\circ$.
The direction of flattened disc ($\textbf{\textit{n}}_{\rm disc}$) gradually changes during the gravitational contraction, from the magnetic pseudo-disc to the rotationally-supported disc.
The transition occurs around $n_{\rm H} \sim 10^{11}\,\cc$ where the B-field strength weakens via the magnetic dissipation (`dead-zone') and the disc wind is divided into two components, `(low-velocity) outflow' and `jet'.
The intermittent accretion on to the central protostar drives knotty mass ejections (`knot').
}
\label{f6}
\end{figure}

Figure~\ref{f6} presents an overview of the simulation result.
The contraction direction of the prestellar core gradually is changed on different scales.
At the early phase in low-dense outer region, B-field is tightly coupled to the collapsing gas through the flux freezing.
The prestellar core contracts along the B-field lines ($\textbf{\textit{B}}_0$) and forms the pseudo-disc supported by both the magnetic drag and the centrifugal force.
At the later phase in high-dense inner region, on the other hand, the magnetic dissipation weakens the B-field and the rotational disc forms around the angular momentum vector ($\textbf{\textit{J}}_0$).
The B-field lines are dragged by the rotationally supported gas and strongly twisted near the small-sized disc at the collapse centre.
As a result, the B-field direction  strongly depends on the scales of the averaged volumes (Fig.~\ref{f3}a).
Then the B-field structure strongly depends on viewing scales: hourglass, tied tube, and spiral pattern at the core-, intermediate-, and disc-scales, respectively (Fig.~\ref{f2}).
Various structures of B-field can be easily reproduced around the same protostar.

Although we could not calculate the evolution exceeding $\sim\!300$\,yr due to the extremely high CPU cost, we could unveil the early phase of the star formation.
To validate the simulation results, high-spatial resolution observations are necessary.
Now, the spatial resolution of ALMA telescope is less than $\sim\!5$\,au in nearby star forming regions \citep{ALMA15}.
Future telescopes would grow in performance.
Thus, the spatial resolution of $\ll\!1$--$5$\,au is necessary to compare simulations with observations.
With the combination of simulations with observations, we would understand the star formation process more precisely.
The spatial resolution of our simulation is $0.01$\,au (Section~\ref{sec:method_ics}).
Thus, our simulation can be compared with recent high-spatial resolution ALMA observations, as described in Sections~\ref{sec:discus} and \ref{sec:discussion}.

Moreover, the high-speed flows (jet and knots) episodically appear near the protostar, which is caused by the gravitationally instability of a massive circumstellar disk, indicating that the mass accretion onto the protostar intermittently occurs.
The time-variable accretion would change the protostellar luminosity in a short duration, which would be confirmed by future observation monitoring the protostellar luminosities.
In addition, understanding the disk properties is crucially important to investigate the planet formation.
Thus, our study can give a great impact for understanding the star formation process.

\section*{Acknowledgements}

The authors would like to thank Hideyuki Kitta, Basmah Riaz, Kengo Tomida, and Kohji Tomisaka for the stimulating discussions.
The present research used the computational resources of the HPCI system provided by (Cyber Sciencecenter, Tohoku University; Cybermedia Center, Osaka University through the HPCI System Research Project (Project ID: hp170047, hp180001). Simulations reported in this paper were also performed by 2017 and 2018 Koubo Kadai on Earth Simulator (NEC SX-ACE) at JAMSTEC.
This work was supported by JSPS Research Fellow to SH and JSPS KAKENHI Grant Numbers 18J01296 to SH and by 17K05387, 17H06360, and 17H02869 to MNM.


\bibliographystyle{mnras}
\bibliography{ms.bbl}

\begin{thebibliography}{}
\makeatletter
\relax
\def\mn@urlcharsother{\let\do\@makeother \do\$\do\&\do\#\do\^\do\_\do\%\do\~}
\def\mn@doi{\begingroup\mn@urlcharsother \@ifnextchar [ {\mn@doi@}
  {\mn@doi@[]}}
\def\mn@doi@[#1]#2{\def\@tempa{#1}\ifx\@tempa\@empty \href
  {http://dx.doi.org/#2} {doi:#2}\else \href {http://dx.doi.org/#2} {#1}\fi
  \endgroup}
\def\mn@eprint#1#2{\mn@eprint@#1:#2::\@nil}
\def\mn@eprint@arXiv#1{\href {http://arxiv.org/abs/#1} {{\tt arXiv:#1}}}
\def\mn@eprint@dblp#1{\href {http://dblp.uni-trier.de/rec/bibtex/#1.xml}
  {dblp:#1}}
\def\mn@eprint@#1:#2:#3:#4\@nil{\def\@tempa {#1}\def\@tempb {#2}\def\@tempc
  {#3}\ifx \@tempc \@empty \let \@tempc \@tempb \let \@tempb \@tempa \fi \ifx
  \@tempb \@empty \def\@tempb {arXiv}\fi \@ifundefined
  {mn@eprint@\@tempb}{\@tempb:\@tempc}{\expandafter \expandafter \csname
  mn@eprint@\@tempb\endcsname \expandafter{\@tempc}}}

\bibitem[\protect\citeauthoryear{{ALMA Partnership} et~al.,}{{ALMA Partnership}
  et~al.}{2015}]{ALMA15}
{ALMA Partnership} et~al., 2015, \mn@doi [\apjl] {10.1088/2041-8205/808/1/L3},
  \href {http://adsabs.harvard.edu/abs/2015ApJ...808L...3A} {808, L3}

\bibitem[\protect\citeauthoryear{{Alves}, {Girart}, {Caselli}, {Franco},
  {Zhao}, {Vlemmings}, {Evans}  \& {Ricci}}{{Alves} et~al.}{2017}]{alves17}
{Alves} F.~O.,  {Girart} J.~M.,  {Caselli} P.,  {Franco} G.~A.~P.,  {Zhao} B.,
  {Vlemmings} W.~H.~T.,  {Evans} M.~G.,   {Ricci} L.,  2017, \mn@doi [\aap]
  {10.1051/0004-6361/201731077}, \href
  {http://adsabs.harvard.edu/abs/2017A%26A...603L...3A} {603, L3}

\bibitem[\protect\citeauthoryear{{Alves} et~al.,}{{Alves}
  et~al.}{2018}]{alves18}
{Alves} F.~O.,  et~al., 2018, \mn@doi [\aap] {10.1051/0004-6361/201832935},
  \href {http://adsabs.harvard.edu/abs/2018A%26A...616A..56A} {616, A56}

\bibitem[\protect\citeauthoryear{{Arce}, {Shepherd}, {Gueth}, {Lee},
  {Bachiller}, {Rosen}  \& {Beuther}}{{Arce} et~al.}{2007}]{arce07}
{Arce} H.~G.,  {Shepherd} D.,  {Gueth} F.,  {Lee} C.-F.,  {Bachiller} R.,
  {Rosen} A.,   {Beuther} H.,  2007, Protostars and Planets V, \href
  {http://adsabs.harvard.edu/abs/2007prpl.conf..245A} {pp 245--260}

\bibitem[\protect\citeauthoryear{{Banerjee} \& {Pudritz}}{{Banerjee} \&
  {Pudritz}}{2006}]{banerjee06}
{Banerjee} R.,  {Pudritz} R.~E.,  2006, \mn@doi [\apj] {10.1086/500496}, \href
  {http://adsabs.harvard.edu/abs/2006ApJ...641..949B} {641, 949}

\bibitem[\protect\citeauthoryear{{Bate}}{{Bate}}{1998}]{bate98}
{Bate} M.~R.,  1998, \mn@doi [\apjl] {10.1086/311719}, \href
  {http://adsabs.harvard.edu/abs/1998ApJ...508L..95B} {508, L95}

\bibitem[\protect\citeauthoryear{{Bjerkeli}, {van der Wiel}, {Harsono},
  {Ramsey}  \& {J{\o}rgensen}}{{Bjerkeli} et~al.}{2016}]{bjerkeli16}
{Bjerkeli} P.,  {van der Wiel} M.~H.~D.,  {Harsono} D.,  {Ramsey} J.~P.,
  {J{\o}rgensen} J.~K.,  2016, \mn@doi [\nat] {10.1038/nature20600}, \href
  {http://adsabs.harvard.edu/abs/2016Natur.540..406B} {540, 406}

\bibitem[\protect\citeauthoryear{{Blandford} \& {Payne}}{{Blandford} \&
  {Payne}}{1982}]{blandford82}
{Blandford} R.~D.,  {Payne} D.~G.,  1982, \mn@doi [\mnras]
  {10.1093/mnras/199.4.883}, \href
  {http://adsabs.harvard.edu/abs/1982MNRAS.199..883B} {199, 883}

\bibitem[\protect\citeauthoryear{{Chapman} et~al.,}{{Chapman}
  et~al.}{2013}]{chapman13}
{Chapman} N.~L.,  et~al., 2013, \mn@doi [\apj] {10.1088/0004-637X/770/2/151},
  \href {http://adsabs.harvard.edu/abs/2013ApJ...770..151C} {770, 151}

\bibitem[\protect\citeauthoryear{{Ciardi} \& {Hennebelle}}{{Ciardi} \&
  {Hennebelle}}{2010}]{ciardi10}
{Ciardi} A.,  {Hennebelle} P.,  2010, \mn@doi [\mnras]
  {10.1111/j.1745-3933.2010.00942.x}, \href
  {http://adsabs.harvard.edu/abs/2010MNRAS.409L..39C} {409, L39}

\bibitem[\protect\citeauthoryear{{Fukuda} \& {Hanawa}}{{Fukuda} \&
  {Hanawa}}{1999}]{fukuda99}
{Fukuda} N.,  {Hanawa} T.,  1999, \mn@doi [\apj] {10.1086/307169}, \href
  {http://adsabs.harvard.edu/abs/1999ApJ...517..226F} {517, 226}

\bibitem[\protect\citeauthoryear{{Galametz} et~al.,}{{Galametz}
  et~al.}{2018}]{galametz18}
{Galametz} M.,  et~al., 2018, \mn@doi [\aap] {10.1051/0004-6361/201833004},
  \href {http://adsabs.harvard.edu/abs/2018A%26A...616A.139G} {616, A139}

\bibitem[\protect\citeauthoryear{{Girart}, {Rao}  \& {Marrone}}{{Girart}
  et~al.}{2006}]{girart06}
{Girart} J.~M.,  {Rao} R.,   {Marrone} D.~P.,  2006, \mn@doi [Science]
  {10.1126/science.1129093}, \href
  {http://adsabs.harvard.edu/abs/2006Sci...313..812G} {313, 812}

\bibitem[\protect\citeauthoryear{{Haro}}{{Haro}}{1952}]{haro50}
{Haro} G.,  1952, \mn@doi [\apj] {10.1086/145576}, \href
  {http://adsabs.harvard.edu/abs/1952ApJ...115..572H} {115, 572}

\bibitem[\protect\citeauthoryear{{Hennebelle} \& {Ciardi}}{{Hennebelle} \&
  {Ciardi}}{2009}]{hennebelle09}
{Hennebelle} P.,  {Ciardi} A.,  2009, \mn@doi [\aap]
  {10.1051/0004-6361/200913008}, \href
  {http://adsabs.harvard.edu/abs/2009A%26A...506L..29H} {506, L29}

\bibitem[\protect\citeauthoryear{{Hennebelle} \& {Fromang}}{{Hennebelle} \&
  {Fromang}}{2008}]{hennebelle08a}
{Hennebelle} P.,  {Fromang} S.,  2008, \mn@doi [\aap]
  {10.1051/0004-6361:20078309}, \href
  {http://adsabs.harvard.edu/abs/2008A%26A...477....9H} {477, 9}

\bibitem[\protect\citeauthoryear{{Hennebelle} \& {Teyssier}}{{Hennebelle} \&
  {Teyssier}}{2008}]{hennebelle08b}
{Hennebelle} P.,  {Teyssier} R.,  2008, \mn@doi [\aap]
  {10.1051/0004-6361:20078310}, \href
  {http://adsabs.harvard.edu/abs/2008A%26A...477...25H} {477, 25}

\bibitem[\protect\citeauthoryear{{Hennebelle}, {Commer{\c c}on}, {Joos},
  {Klessen}, {Krumholz}, {Tan}  \& {Teyssier}}{{Hennebelle}
  et~al.}{2011}]{hennebelle11}
{Hennebelle} P.,  {Commer{\c c}on} B.,  {Joos} M.,  {Klessen} R.~S.,
  {Krumholz} M.,  {Tan} J.~C.,   {Teyssier} R.,  2011, \mn@doi [\aap]
  {10.1051/0004-6361/201016052}, \href
  {http://adsabs.harvard.edu/abs/2011A%26A...528A..72H} {528, A72}

\bibitem[\protect\citeauthoryear{{Hennebelle}, {Commer{\c c}on}, {Chabrier}  \&
  {Marchand}}{{Hennebelle} et~al.}{2016}]{hennebelle16}
{Hennebelle} P.,  {Commer{\c c}on} B.,  {Chabrier} G.,   {Marchand} P.,  2016,
  \mn@doi [\apjl] {10.3847/2041-8205/830/1/L8}, \href
  {http://adsabs.harvard.edu/abs/2016ApJ...830L...8H} {830, L8}

\bibitem[\protect\citeauthoryear{{Herbig}}{{Herbig}}{1951}]{herbig51}
{Herbig} G.~H.,  1951, \mn@doi [\apj] {10.1086/145440}, \href
  {http://adsabs.harvard.edu/abs/1951ApJ...113..697H} {113, 697}

\bibitem[\protect\citeauthoryear{{Hirota}, {Machida}, {Matsushita}, {Motogi},
  {Matsumoto}, {Kim}, {Burns}  \& {Honma}}{{Hirota} et~al.}{2017}]{hirota17}
{Hirota} T.,  {Machida} M.~N.,  {Matsushita} Y.,  {Motogi} K.,  {Matsumoto} N.,
   {Kim} M.~K.,  {Burns} R.~A.,   {Honma} M.,  2017, \mn@doi [Nature Astronomy]
  {10.1038/s41550-017-0146}, \href
  {http://adsabs.harvard.edu/abs/2017NatAs...1E.146H} {1, 0146}

\bibitem[\protect\citeauthoryear{{Hull} et~al.,}{{Hull} et~al.}{2013}]{hull13}
{Hull} C.~L.~H.,  et~al., 2013, \mn@doi [\apj] {10.1088/0004-637X/768/2/159},
  \href {http://adsabs.harvard.edu/abs/2013ApJ...768..159H} {768, 159}

\bibitem[\protect\citeauthoryear{{Hull} et~al.,}{{Hull} et~al.}{2017}]{hull17}
{Hull} C.~L.~H.,  et~al., 2017, \mn@doi [\apj] {10.3847/1538-4357/aa7fe9},
  \href {http://adsabs.harvard.edu/abs/2017ApJ...847...92H} {847, 92}

\bibitem[\protect\citeauthoryear{{Joos}, {Hennebelle}  \& {Ciardi}}{{Joos}
  et~al.}{2012}]{joos12}
{Joos} M.,  {Hennebelle} P.,   {Ciardi} A.,  2012, \mn@doi [\aap]
  {10.1051/0004-6361/201118730}, \href
  {http://adsabs.harvard.edu/abs/2012A%26A...543A.128J} {543, A128}

\bibitem[\protect\citeauthoryear{{Konigl} \& {Pudritz}}{{Konigl} \&
  {Pudritz}}{2000}]{konigl00}
{Konigl} A.,  {Pudritz} R.~E.,  2000, Protostars and Planets IV, \href
  {http://adsabs.harvard.edu/abs/2000prpl.conf..759K} {p.~759}

\bibitem[\protect\citeauthoryear{{Larson}}{{Larson}}{2003}]{larson03}
{Larson} R.~B.,  2003, \mn@doi [Reports on Progress in Physics]
  {10.1088/0034-4885/66/10/R03}, \href
  {http://adsabs.harvard.edu/abs/2003RPPh...66.1651L} {66, 1651}

\bibitem[\protect\citeauthoryear{{Lee}, {Ho}, {Li}, {Hirano}, {Zhang}  \&
  {Shang}}{{Lee} et~al.}{2017}]{lee17}
{Lee} C.-F.,  {Ho} P. T.~P.,  {Li} Z.-Y.,  {Hirano} N.,  {Zhang} Q.,   {Shang}
  H.,  2017, \mn@doi [Nature Astronomy] {10.1038/s41550-017-0152}, \href
  {https://ui.adsabs.harvard.edu/#abs/2017NatAs...1E.152L} {1, 0152}

\bibitem[\protect\citeauthoryear{{Lee}, {Li}, {Hirano}, {Shang}, {Ho}  \&
  {Zhang}}{{Lee} et~al.}{2018}]{lee18}
{Lee} C.-F.,  {Li} Z.-Y.,  {Hirano} N.,  {Shang} H.,  {Ho} P.~T.~P.,   {Zhang}
  Q.,  2018, \mn@doi [\apj] {10.3847/1538-4357/aad2da}, \href
  {http://adsabs.harvard.edu/abs/2018ApJ...863...94L} {863, 94}

\bibitem[\protect\citeauthoryear{{Lewis} \& {Bate}}{{Lewis} \&
  {Bate}}{2018}]{lewis18}
{Lewis} B.~T.,  {Bate} M.~R.,  2018, \mn@doi [\mnras] {10.1093/mnras/sty829},
  \href {http://adsabs.harvard.edu/abs/2018MNRAS.477.4241L} {477, 4241}

\bibitem[\protect\citeauthoryear{{Li}, {Banerjee}, {Pudritz}, {J{\o}rgensen},
  {Shang}, {Krasnopolsky}  \& {Maury}}{{Li} et~al.}{2014}]{li14}
{Li} Z.-Y.,  {Banerjee} R.,  {Pudritz} R.~E.,  {J{\o}rgensen} J.~K.,  {Shang}
  H.,  {Krasnopolsky} R.,   {Maury} A.,  2014, \mn@doi [Protostars and Planets
  VI] {10.2458/azu_uapress_9780816531240-ch008}, \href
  {http://adsabs.harvard.edu/abs/2014prpl.conf..173L} {pp 173--194}

\bibitem[\protect\citeauthoryear{{Lynden-Bell}}{{Lynden-Bell}}{2003}]{lynden-bell03}
{Lynden-Bell} D.,  2003, \mn@doi [\mnras] {10.1046/j.1365-8711.2003.06506.x},
  \href {http://adsabs.harvard.edu/abs/2003MNRAS.341.1360L} {341, 1360}

\bibitem[\protect\citeauthoryear{{Machida}}{{Machida}}{2014}]{machida14}
{Machida} M.~N.,  2014, \mn@doi [\apjl] {10.1088/2041-8205/796/1/L17}, \href
  {http://adsabs.harvard.edu/abs/2014ApJ...796L..17M} {796, L17}

\bibitem[\protect\citeauthoryear{{Machida} \& {Matsumoto}}{{Machida} \&
  {Matsumoto}}{2011}]{machida11}
{Machida} M.~N.,  {Matsumoto} T.,  2011, \mn@doi [\mnras]
  {10.1111/j.1365-2966.2011.18349.x}, \href
  {http://adsabs.harvard.edu/abs/2011MNRAS.413.2767M} {413, 2767}

\bibitem[\protect\citeauthoryear{{Machida}, {Matsumoto}, {Hanawa}  \&
  {Tomisaka}}{{Machida} et~al.}{2006}]{machida06}
{Machida} M.~N.,  {Matsumoto} T.,  {Hanawa} T.,   {Tomisaka} K.,  2006, \mn@doi
  [\apj] {10.1086/504423}, \href
  {http://adsabs.harvard.edu/abs/2006ApJ...645.1227M} {645, 1227}

\bibitem[\protect\citeauthoryear{{Machida}, {Inutsuka}  \&
  {Matsumoto}}{{Machida} et~al.}{2007}]{machida07}
{Machida} M.~N.,  {Inutsuka} S.-i.,   {Matsumoto} T.,  2007, \mn@doi [\apj]
  {10.1086/521779}, \href {http://adsabs.harvard.edu/abs/2007ApJ...670.1198M}
  {670, 1198}

\bibitem[\protect\citeauthoryear{{Machida}, {Inutsuka}  \&
  {Matsumoto}}{{Machida} et~al.}{2008}]{machida08}
{Machida} M.~N.,  {Inutsuka} S.-i.,   {Matsumoto} T.,  2008, \mn@doi [\apj]
  {10.1086/528364}, \href {http://adsabs.harvard.edu/abs/2008ApJ...676.1088M}
  {676, 1088}

\bibitem[\protect\citeauthoryear{{Machida}, {Inutsuka}  \&
  {Matsumoto}}{{Machida} et~al.}{2010}]{machida10}
{Machida} M.~N.,  {Inutsuka} S.-i.,   {Matsumoto} T.,  2010, \mn@doi [\apj]
  {10.1088/0004-637X/724/2/1006}, \href
  {http://adsabs.harvard.edu/abs/2010ApJ...724.1006M} {724, 1006}

\bibitem[\protect\citeauthoryear{{Machida}, {Inutsuka}  \&
  {Matsumoto}}{{Machida} et~al.}{2014}]{machida14a}
{Machida} M.~N.,  {Inutsuka} S.-i.,   {Matsumoto} T.,  2014, \mn@doi [\mnras]
  {10.1093/mnras/stt2343}, \href
  {http://adsabs.harvard.edu/abs/2014MNRAS.438.2278M} {438, 2278}

\bibitem[\protect\citeauthoryear{{Marchand}, {Masson}, {Chabrier},
  {Hennebelle}, {Commer{\c c}on}  \& {Vaytet}}{{Marchand}
  et~al.}{2016}]{marchand16}
{Marchand} P.,  {Masson} J.,  {Chabrier} G.,  {Hennebelle} P.,  {Commer{\c
  c}on} B.,   {Vaytet} N.,  2016, \mn@doi [\aap] {10.1051/0004-6361/201526780},
  \href {http://adsabs.harvard.edu/abs/2016A%26A...592A..18M} {592, A18}

\bibitem[\protect\citeauthoryear{{Matsumoto} \& {Hanawa}}{{Matsumoto} \&
  {Hanawa}}{2003}]{matsumoto03}
{Matsumoto} T.,  {Hanawa} T.,  2003, \mn@doi [\apj] {10.1086/345338}, \href
  {http://adsabs.harvard.edu/abs/2003ApJ...583..296M} {583, 296}

\bibitem[\protect\citeauthoryear{{Matsumoto} \& {Tomisaka}}{{Matsumoto} \&
  {Tomisaka}}{2004}]{matsumoto04}
{Matsumoto} T.,  {Tomisaka} K.,  2004, \mn@doi [\apj] {10.1086/424897}, \href
  {http://adsabs.harvard.edu/abs/2004ApJ...616..266M} {616, 266}

\bibitem[\protect\citeauthoryear{{Matsumoto}, {Machida}  \&
  {Inutsuka}}{{Matsumoto} et~al.}{2017}]{matsumoto17}
{Matsumoto} T.,  {Machida} M.~N.,   {Inutsuka} S.-i.,  2017, \mn@doi [\apj]
  {10.3847/1538-4357/aa6a1c}, \href
  {http://adsabs.harvard.edu/abs/2017ApJ...839...69M} {839, 69}

\bibitem[\protect\citeauthoryear{{Matsushita}, {Takahashi}, {Machida}  \&
  {Tomisaka}}{{Matsushita} et~al.}{2019}]{matsushita19}
{Matsushita} Y.,  {Takahashi} S.,  {Machida} M.~N.,   {Tomisaka} K.,  2019,
  \mn@doi [\apj] {10.3847/1538-4357/aaf1b6}, \href
  {http://adsabs.harvard.edu/abs/2019ApJ...871..221M} {871, 221}

\bibitem[\protect\citeauthoryear{{Mayama} et~al.,}{{Mayama}
  et~al.}{2018}]{mayama18}
{Mayama} S.,  et~al., 2018, \mn@doi [\apjl] {10.3847/2041-8213/aae88b}, \href
  {http://adsabs.harvard.edu/abs/2018ApJ...868L...3M} {868, L3}

\bibitem[\protect\citeauthoryear{{Nakano}, {Nishi}  \& {Umebayashi}}{{Nakano}
  et~al.}{2002}]{nakano02}
{Nakano} T.,  {Nishi} R.,   {Umebayashi} T.,  2002, \mn@doi [\apj]
  {10.1086/340587}, \href {http://adsabs.harvard.edu/abs/2002ApJ...573..199N}
  {573, 199}

\bibitem[\protect\citeauthoryear{{Okoda}, {Oya}, {Sakai}, {Watanabe},
  {J{\o}rgensen}, {Van Dishoeck}  \& {Yamamoto}}{{Okoda}
  et~al.}{2018}]{okoda18}
{Okoda} Y.,  {Oya} Y.,  {Sakai} N.,  {Watanabe} Y.,  {J{\o}rgensen} J.~K.,
  {Van Dishoeck} E.~F.,   {Yamamoto} S.,  2018, \mn@doi [\apjl]
  {10.3847/2041-8213/aad8ba}, \href
  {http://adsabs.harvard.edu/abs/2018ApJ...864L..25O} {864, L25}

\bibitem[\protect\citeauthoryear{{Pudritz}, {Ouyed}, {Fendt}  \&
  {Brandenburg}}{{Pudritz} et~al.}{2007}]{pudritz07}
{Pudritz} R.~E.,  {Ouyed} R.,  {Fendt} C.,   {Brandenburg} A.,  2007,
  Protostars and Planets V, \href
  {http://adsabs.harvard.edu/abs/2007prpl.conf..277P} {pp 277--294}

\bibitem[\protect\citeauthoryear{{Reipurth} \& {Bally}}{{Reipurth} \&
  {Bally}}{2001}]{reipurth01}
{Reipurth} B.,  {Bally} J.,  2001, \mn@doi [\araa]
  {10.1146/annurev.astro.39.1.403}, \href
  {http://adsabs.harvard.edu/abs/2001ARA\%26A..39..403R} {39, 403}

\bibitem[\protect\citeauthoryear{{Riaz}, {Brice{\~n}o}, {Whelan}  \&
  {Heathcote}}{{Riaz} et~al.}{2017}]{riaz17}
{Riaz} B.,  {Brice{\~n}o} C.,  {Whelan} E.~T.,   {Heathcote} S.,  2017, \mn@doi
  [\apj] {10.3847/1538-4357/aa70e8}, \href
  {http://adsabs.harvard.edu/abs/2017ApJ...844...47R} {844, 47}

\bibitem[\protect\citeauthoryear{{Sakai}, {Hanawa}, {Zhang}, {Higuchi},
  {Ohashi}, {Oya}  \& &~{Yamamoto}}{{Sakai} et~al.}{2019}]{sakai19}
{Sakai} N.,  {Hanawa} T.,  {Zhang} Y.,  {Higuchi} A.~E.,  {Ohashi} S.,  {Oya}
  Y.,  {Yamamoto} S.,  2019, \mn@doi [\nat] {10.1038/nature20600}, \href
  {https://www.nature.com/articles/s41586-018-0819-2?WT.feed_name=subjects_astronomy-and-astrophysics}
  {565, 206}

\bibitem[\protect\citeauthoryear{{Seifried}, {Banerjee}, {Pudritz}  \&
  {Klessen}}{{Seifried} et~al.}{2012}]{seifried12}
{Seifried} D.,  {Banerjee} R.,  {Pudritz} R.~E.,   {Klessen} R.~S.,  2012,
  \mn@doi [\mnras] {10.1111/j.1745-3933.2012.01253.x}, \href
  {http://adsabs.harvard.edu/abs/2012MNRAS.423L..40S} {423, L40}

\bibitem[\protect\citeauthoryear{{Seifried}, {Banerjee}, {Pudritz}  \&
  {Klessen}}{{Seifried} et~al.}{2013}]{seifried13}
{Seifried} D.,  {Banerjee} R.,  {Pudritz} R.~E.,   {Klessen} R.~S.,  2013,
  \mn@doi [\mnras] {10.1093/mnras/stt682}, \href
  {http://adsabs.harvard.edu/abs/2013MNRAS.432.3320S} {432, 3320}

\bibitem[\protect\citeauthoryear{{Seifried}, {Banerjee}, {Pudritz}  \&
  {Klessen}}{{Seifried} et~al.}{2015}]{seifried15}
{Seifried} D.,  {Banerjee} R.,  {Pudritz} R.~E.,   {Klessen} R.~S.,  2015,
  \mn@doi [\mnras] {10.1093/mnras/stu2282}, \href
  {http://adsabs.harvard.edu/abs/2015MNRAS.446.2776S} {446, 2776}

\bibitem[\protect\citeauthoryear{{Shinnaga} et~al.,}{{Shinnaga}
  et~al.}{2012}]{shinnaga12}
{Shinnaga} H.,  et~al., 2012, \mn@doi [\apjl] {10.1088/2041-8205/750/2/L29},
  \href {http://adsabs.harvard.edu/abs/2012ApJ...750L..29S} {750, L29}

\bibitem[\protect\citeauthoryear{{Shu}}{{Shu}}{1977}]{shu77}
{Shu} F.~H.,  1977, \mn@doi [\apj] {10.1086/155274}, \href
  {http://adsabs.harvard.edu/abs/1977ApJ...214..488S} {214, 488}

\bibitem[\protect\citeauthoryear{{Stephens} et~al.,}{{Stephens}
  et~al.}{2017}]{stephens17}
{Stephens} I.~W.,  et~al., 2017, \mn@doi [\apj] {10.3847/1538-4357/aa8262},
  \href {http://adsabs.harvard.edu/abs/2017ApJ...846...16S} {846, 16}

\bibitem[\protect\citeauthoryear{{Takami}, {Karr}, {Nisini}  \& {Ray}}{{Takami}
  et~al.}{2011}]{takami11}
{Takami} M.,  {Karr} J.~L.,  {Nisini} B.,   {Ray} T.~P.,  2011, \mn@doi [\apj]
  {10.1088/0004-637X/743/2/193}, \href
  {http://adsabs.harvard.edu/abs/2011ApJ...743..193T} {743, 193}

\bibitem[\protect\citeauthoryear{{Tomida}, {Tomisaka}, {Matsumoto}, {Hori},
  {Okuzumi}, {Machida}  \& {Saigo}}{{Tomida} et~al.}{2013}]{tomida13}
{Tomida} K.,  {Tomisaka} K.,  {Matsumoto} T.,  {Hori} Y.,  {Okuzumi} S.,
  {Machida} M.~N.,   {Saigo} K.,  2013, \mn@doi [\apj]
  {10.1088/0004-637X/763/1/6}, \href
  {http://adsabs.harvard.edu/abs/2013ApJ...763....6T} {763, 6}

\bibitem[\protect\citeauthoryear{{Tomida}, {Okuzumi}  \& {Machida}}{{Tomida}
  et~al.}{2015}]{tomida15}
{Tomida} K.,  {Okuzumi} S.,   {Machida} M.~N.,  2015, \mn@doi [\apj]
  {10.1088/0004-637X/801/2/117}, \href
  {http://adsabs.harvard.edu/abs/2015ApJ...801..117T} {801, 117}

\bibitem[\protect\citeauthoryear{{Tomisaka}}{{Tomisaka}}{2002}]{tomisaka02}
{Tomisaka} K.,  2002, \mn@doi [\apj] {10.1086/341133}, \href
  {http://adsabs.harvard.edu/abs/2002ApJ...575..306T} {575, 306}

\bibitem[\protect\citeauthoryear{{Tsukamoto}, {Iwasaki}, {Okuzumi}, {Machida}
  \& {Inutsuka}}{{Tsukamoto} et~al.}{2015}]{tsukamoto15}
{Tsukamoto} Y.,  {Iwasaki} K.,  {Okuzumi} S.,  {Machida} M.~N.,   {Inutsuka}
  S.,  2015, \mn@doi [\mnras] {10.1093/mnras/stv1290}, \href
  {http://ads.nao.ac.jp/abs/2015MNRAS.452..278T} {452, 278}

\bibitem[\protect\citeauthoryear{{Wurster}, {Bate}  \& {Price}}{{Wurster}
  et~al.}{2018}]{wurster18}
{Wurster} J.,  {Bate} M.~R.,   {Price} D.~J.,  2018, \mn@doi [\mnras]
  {10.1093/mnras/sty2438}, \href
  {http://adsabs.harvard.edu/abs/2018MNRAS.481.2450W} {481, 2450}

\makeatother
\end{thebibliography}

\bsp	
\label{lastpage}
\end{document}